\documentclass[11pt]{article}
\pdfoutput=1
\usepackage[utf8]{inputenc}
\usepackage{multirow}
\usepackage{svg}
\usepackage{amsmath, amsfonts, amssymb}
\usepackage{comment}
\usepackage{graphicx}
\usepackage{psfrag}
\usepackage{amsthm}
\usepackage{xcolor}
\usepackage{enumerate}
\usepackage{arydshln}
\usepackage{soul}
 \usepackage{slashed}
\usepackage{subfig}
\usepackage{mathrsfs}
 \usepackage{a4wide}
  \usepackage{tikz}
  \usepackage{tikz-cd}
  \usetikzlibrary{shapes.geometric}
  \usepackage{tcolorbox}
\usepackage{wrapfig}
  \usepackage{color}
  \definecolor{dark-gray}{gray}{0.20}
  \definecolor{gray}{gray}{0.30}
  \definecolor{light-gray}{gray}{0.80}
  \definecolor{dark-red}{rgb}{0.7,0,0}
  \definecolor{dark-green}{rgb}{0.1,0.4,0}
  \definecolor{dark-blue}{rgb}{0.3,0.3,0.7}
  \definecolor{light-blue}{rgb}{0.8,0.8,1}
      \definecolor{swamp}{RGB}{240, 199, 197}

  \usepackage{pifont}

\usepackage{newunicodechar} 

\usepackage{setspace}

\newcommand{\be}{\begin{equation}}
\newcommand{\ee}{\end{equation}}

\def\be{\begin{equation}}
\def\ee{\end{equation}}
\def\bea{\begin{eqnarray}}
\def\eea{\end{eqnarray}}

\definecolor{antiquefuchsia}{rgb}{0.57, 0.36, 0.51}

\def\simleq{\; \raise0.3ex\hbox{$<$\kern-0.75em
      \raise-1.1ex\hbox{$\sim$}}\; }
   \def\simgeq{\; \raise0.3ex\hbox{$>$\kern-0.75em
      \raise-1.1ex\hbox{$\sim$}}\; }

\numberwithin{equation}{section}

\usepackage{jheppub}
\usepackage{hyperref}
\usepackage{cleveref}

\hypersetup{
	colorlinks=true,
	linkcolor=antiquefuchsia,
	citecolor=antiquefuchsia,
	urlcolor=antiquefuchsia,
	linktoc=page
}

\theoremstyle{remark}

\crefname{appendix}{Appendix}{Appendices}

\title{\centering  The Bubble of Nothing under T-duality}

\author{Matilda Delgado$^1$} \affiliation{$^1$Instituto de F\'{i}sica Te\'{o}rica IFT-UAM/CSIC,
C/ Nicol\'{a}s Cabrera 13-15, Campus de Cantoblanco, 28049 Madrid, Spain}

\emailAdd{matilda.delgado@uam.es}

\abstract{The bubble of nothing is a solution to Einstein's equations where a circle shrinks and pinches off smoothly. As such, it is one of the simplest examples of a dynamical cobordism to nothing. We take a first step in studying how this solution transforms under T-duality in bosonic string theory. Applying Buscher's rules reveals that the dual solution features a singular, strongly coupled core, with a circle blowing-up rather than pinching off. This naive approach to T-duality solely accounts for the zero-modes of the fields after dimensional reduction on the circle. For this reason, we argue that this is not the full picture that the T-dual solution should depend non-trivially on the dual circle. We point out evidence to this effect both in the gravity description and on the worldsheet. A more complete description of the T-dual object would require a full-fledged sigma model for the bubble of nothing. Nevertheless, inspired by similar examples in the literature, we detail one possible scenario where the stringy bubble of nothing is mediated by closed string tachyon condensation and we discuss its T-duality. }

\begin{document}
\hypersetup{pageanchor=false}
\makeatletter
\let\old@fpheader\@fpheader
\preprint{IFT-UAM/CSIC-23-146}

\makeatother

\maketitle

\hypersetup{
    pdftitle={Title},
    pdfauthor={Matilda Delgado},
    pdfsubject={String dualities, Cobordisms}
}

\section{Introduction and Summary}

The bubble of nothing (BoN), first constructed by Witten \cite{Witten:1981gj}, describes an instability of the Kaluza-Klein vacuum, where the compact circle collapses into a point. This instability is mediated by a bubble that expands at the speed of light, consuming all of spacetime and leaving ``nothing" behind. The BoN is thus the prototypical example of vacuum decay into nothing. Being a simple solution to Einstein's equations in the vacuum, it has been generalized (see e.g. \cite{Yang:2009wz,Blanco-Pillado:2010xww,Blanco-Pillado:2023aom,Friedrich:2023tid}) and studied in various contexts. In recent years, it has played a prominent role in understanding the stability of non-supersymmetric vacua in string theory \cite{Blanco-Pillado:2016xvf,Ooguri:2017njy,GarciaEtxebarria:2020xsr, Dibitetto:2020csn, Bomans:2021ara, Petri:2022yhy}. It has also generated interest for its implications in the construction of de Sitter vacua in string theory \cite{Draper:2021qtc, Draper:2021ujg}.  Finally, as a configuration that ends spacetime, it naturally aligns with the Cobordism Conjecture \cite{McNamara:2019rup} (CC) in the context of the Swampland Program \cite{Vafa:2005ui}.

The CC posits that there should always exist finite-energy configurations that terminate spacetime in any Effective Field Theory (EFT) compatible with Quantum Gravity. It can be viewed as a specific instance of the No Global Symmetries Conjecture \cite{Banks:2010zn, Harlow:2018tng, Harlow:2018jwu}, which asserts that all global symmetries in quantum gravity should be gauged or broken. The CC specifically addresses global symmetries arising from uncancelled topological charges in the compactification manifold. It predicts the existence of corresponding objects, termed ``cobordism defects", that absorb the charge in such a way that the global symmetry is broken and spacetime can terminate smoothly. The CC has led to a focus on identifying these defects and the resulting spacetime-ending configurations, often referred to as ``cobordisms to nothing" (see, for example, \cite{McNamara:2019rup, Dierigl:2022reg, Debray:2023yrs, McNamara:2022lrw}).

Going beyond the CC as a topological statement, one can investigate spacetime-ending configurations by engineering setups in which the compact space dynamically pinches off in spacetime. In other words, one can consider solutions to the (super-)gravity equations of motion that evolves along a spacetime dimension in such a way that the compact space shrinks to a point dynamically. These solutions describing dynamical cobordisms to nothing were introduced in \cite{Buratti:2021fiv} and further explored in \cite{Angius:2022aeq} (see also \cite{Blumenhagen:2022mqw, Angius:2022mgh,Blumenhagen:2022bvh,Basile:2022ypo, Blumenhagen:2023abk})\footnote{For the related topic of solutions in theories with dynamical tadpoles, see \cite{Dudas:2000ff,Blumenhagen:2000dc,Dudas:2002dg,Dudas:2004nd} for early work and \cite{Buratti:2021yia,Raucci:2022jgw,Antonelli:2019nar,Mininno:2020sdb,Basile:2020xwi,Basile:2021mkd,Mourad:2022loy} for related recent developments.}. The BoN serves as one of the simplest examples of such dynamical realizations of cobordisms to nothing (for related work on the bubble of nothing in the context of the cobordism conjecture, see \cite{Friedrich:2023tid}).   \\
\begin{wrapfigure}{l}{0.4\textwidth}
\includegraphics[width=0.9\linewidth]{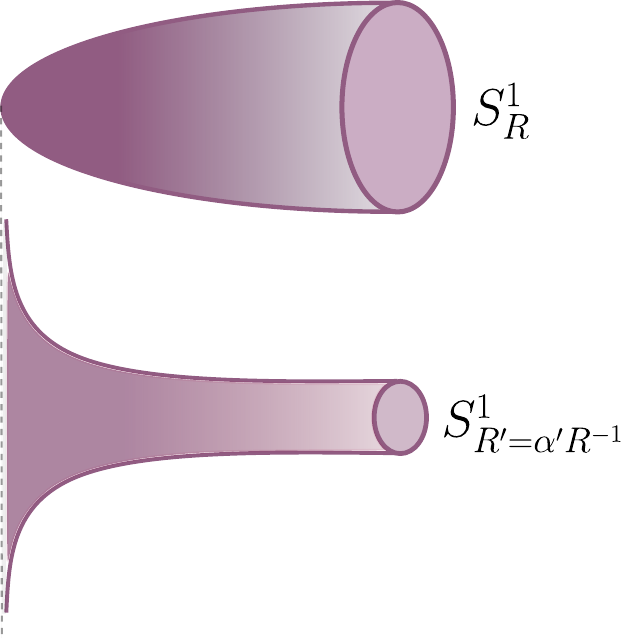} 
\caption{The naive picture of how the bubble of nothing geometry would transform under T-duality. As the radius of the circle pinches off, one would expect the radius of the dual circle to blow up.}
\label{fig2}
\end{wrapfigure}
One of the original motivations for this work was to understand how dynamical realizations of cobordisms to nothing behave under T-duality. T-duality describes the equivalence between the perturbative spectrum of a string theory on a circle of radius R and that of a (possibly different) string theory on a radius $\alpha' R^{-1}$. The intuitive expectation is that the T-dual of a cobordism to nothing with $R\to 0$ should resemble a spontaneous decompactification with $R^{-1}\to \infty$. If the CC asserts that configurations exist where the compact space shrinks to nothing, does T-duality imply the existence of configurations where the compact space expands and decompactifies instead? This is depicted schematically in Figure \ref{fig2}. While this picture appears intuitive, there are fundamental reasons to believe that cobordisms to nothing and spontaneous decompactifications cannot be related by T-duality. On one hand, cobordisms to nothing are meant to be ``finite action" processes, while spontaneous decompactification requires an infinite amount of energy. One can see that a decompactification takes an infinite amount of energy from the fact that an infinite tower of massive KK modes needs to be taken into account in order to account for the new extended dimension(s). This prevents these two types of processes from being related by T-duality: a finite energy state cannot be T-dual to an infinite energy state. Moreover, in cases where T-duality has been employed in the context of the CC, it ended up relating two spacetime-ending configurations and allowed one to obtain one cobordism defect from another. For example, the cobordism defect of Type IIB on a circle was derived by T-dualizing that of Type IIA in \cite{McNamara:2019rup}. This suggests that the intuitive picture of the dual solution describing a spontaneous decompactification is too naive and that the question needs to be studied in more detail.\\

In this note, we aim to shed light on the T-duality of the simplest possible example of a dynamical cobordism to nothing: the BoN in bosonic string theory. The BoN describes a weakly-coupled and smooth configuration that ends spacetime. Furthermore, we consider bosonic string theory because a subtlety arises when fermions are taken into account: fermions possess anti-periodic boundary conditions, creating a topological obstruction to the circle's pinch-off in cases where the circle has periodic boundary conditions. To address this, the CC predicts the existence of a suitable ``spin-defect" at the point of pinch-off. The identification of this spin-defect is case-dependent and generally unknown. To circumvent this subtlety, we thus focus our discussion in this note on cases without fermions, making the BoN topologically allowed without the need for new defects in the configuration. To be concrete, we will pick bosonic string theory since there are no fermions in its the perturbative spectrum. We nevertheless expect our results to extend to more general scenarios. In particular, they naturally extend to any other string theory without fermions. We also expect them to extend to cases with fermions, when the circle is taken with anti-periodic boundary conditions. We will comment on these potential extensions punctually throughout this note.

Examining this straightforward case can help us understand whether the T-dual of a cobordism to nothing can indeed describe spontaneous decompactification. Regardless of the outcome, this analysis will provide insights into this new object in bosonic string theory. Indeed, studying the T-dual solution of the BoN is interesting in of itself and has, to our knowledge, not been done in the literature. Furthermore, our discussion will lead us to glimpsing at the worldsheet description of the BoN and its dual. This will bring light to the stringy mechanisms at play in a full-fledged string-theoretic embedding of a cobordism to nothing.

\begin{figure}
    \centering
    \includegraphics[width=0.5\textwidth]{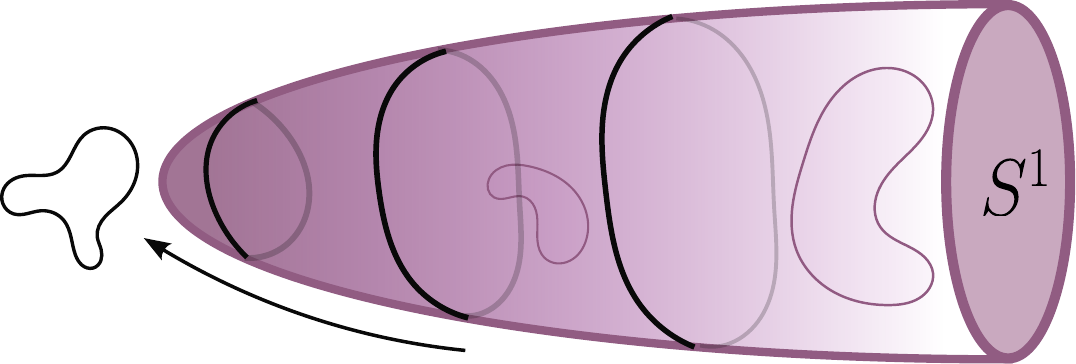}
    \vspace{15pt}
    \caption{A sketch of momentum and winding modes of the string along the circle that pinches off. Modes that wind around the circle can escape by going to the pinch-off point. Winding number is therefore not a conserved quantum number.}
    \label{figure1}
\end{figure}

We will demonstrate that, when using the straightforward approach to T-duality outlined by the Buscher rules, we obtain the intuitive expectation: a solution dependent solely on the radial coordinate, where the size of the circle blows up instead of pinching off. However, we will argue that this approach provides an incomplete picture. To understand how the BoN changes under T-duality, one can consider the conserved U(1) symmetries in bosonic string theory compactified on a circle. These symmetries correspond to momentum states along the circle and winding states around it. In the case of the BoN, since the circle smoothly pinches off, winding states encounter no obstruction to simply sliding off the circle. This is sketched in Figure \ref{figure1}. This implies that while momentum charge is conserved in the BoN, winding charge is not. Under T-duality, momentum and winding states interchange. Consequently, the T-dual of the BoN will not feature conserved momentum charge. This means that the true T-dual solution should depend on the angular coordinate along the dual circle \cite{Gregory:1997te}. The solution provided by the Buscher rules must therefore be but an approximation of the real one. This is not surprising since Buscher rules are only sensitive to the zero modes of the 25-dimensional fields obtained from the dimensional reduction on the circle. This paints a picture where some KK modes in the T-dual frame become light and contribute to the geometry near the singularity such that the solution is modified and is granted a dependence on the dual circle.

This reasoning mirrors the process used to T-dualize the Taub-NUT geometry in \cite{Gregory:1997te,Tong:2002rq} (see also \cite{Kimura:2013zva,Kimura:2018hph}). The Taub-NUT geometry resembles the BoN in that a circle smoothly shrinks to a point. It is well-known that performing a T-duality along this circle results in an NS5 brane localized on the circle \cite{Callan:1991dj}. However, as highlighted in \cite{Tong:2002rq}, the Buscher rules do not fully account for this effect and instead yield a solution dependent only on the radial coordinate, describing a smeared NS5 brane along the circle.

This phenomenon also occurs in the case of the cigar geometry, which describes a circle pinching off in a dilaton background. This geometry has a well-understood world-sheet theory, the so-called ``cigar CFT'' which is a gauged WZW model based on the coset $SL(2,R)/U(1)$ \cite{Witten:1991yr}. The T-duality of this sigma-model is known \cite{Kiritsis:1991zt} and leads to the trumpet CFT which describes a circle that blows up and decompactifies. The locus of this decompactification is singular and at strong coupling. Although the breaking of the winding (resp. momentum) U(1) in the cigar (resp. trumpet) CFT is not immediately seen from the metric and dilaton profile, it becomes obvious when taking the perspective of FZZ duality \cite{FZZ} (see review in \cite{Kazakov:2000pm}). This duality conjectures that the bosonic cigar CFT is equivalent to a Sine-Liouville model (i.e. a $c=1$ coupled to a Liouville field). The breaking of the winding U(1) is explicit in this model due to the presence of a potential. Importantly, one can view the Sine-Liouville potential as the expectation value of a closed string  winding mode. 


We will find many similarities between the T-duality of the cigar and that of the BoN throughout this note. We emphasize, however, that they are two distinct geometries that cannot be deformed into one another. Indeed, in contrast with the cigar geometry, the BoN is Ricci-flat and is as such a smooth, geometric end of spacetime. Furthermore, the BoN is a time-dependent solution, whilst the cigar is static. For these reasons, the methods developed to understand the cigar and the trumpet cannot be directly applied to the BoN.
\\

In this note, we argue that a similar scenario applies to the T-dual of the BoN: the exact T-dual solution depends non-trivially on the dual circle and the solution obtained through the Buscher rules corresponds to an effective smearing of the exact solution along the circle.

To support this claim, we first perform a general relativistic analysis and find indications that the solution produced by the Buscher rules indeed corresponds to a smeared solution. This can be done simply by studying the equations of motion. Then, we provide some evidence that the tension of the dual object is indeed finite. Although we do not have access to the localized solution of the T-dual of the BoN, we can used the smeared solution as a proxy when the circle is compactified, since the angular dependence would get washed out by the compactification anyway. Indeed, any non-trivial dependence on the angular coordinate should be irrelevant in the lower-dimensional picture. In fact, following T-duality, the solution in the lower-dimensional theory should be \textit{exactly the same} as that of the BoN compactified on the circle. We will make this explicit by performing the dimensional reductions of both solutions (the BoN and the smeared dual solution) on the circle. Importantly, we will show that both solutions possess the same, finite tension. This indicates that the T-dual object has finite tension, which is inconsistent with the spontaneous decompactification depicted by the Buscher rules.

We then endeavour to go beyond the gravity description and take a first step in studying this T-duality from the perspective of the worldsheet. In order to describe the T-dual solution with a large radius, that is, with Einstein gravity, we are forced to consider the T-duality of a string-sized BoN. In this limit, winding modes become light and we lose control of the worldsheet theory. Nevertheless, since the winding modes depend on the dual circle, this gives a rationale for why the T-dual solution breaks the circle isometry. Relatedly, we detail how these modes can account for breaking the U(1) winding symmetry. We then outline further steps that need to be taken in order to better understand the worldsheet theory of the BoN (and its dual). Only with a full-fledged sigma-model for the BoN can one hope to elucidate what happens when it becomes string-sized.

Finally, inspired by the works of \cite{FZZ,Adams:2001sv,Adams:2005rb,McGreevy:2005ci,Fabinger:2000jd}, we outline one possibility for a stringy realization of the BoN. It could be that when the circle is string-sized, winding tachyons could deform the worldsheet theory such that the string modes become massive, effectively ending spacetime. Indeed, although closed string tachyon condensation is generally not well understood, it has long been regarded as describing the decay of spacetime itself. In a nutshell, the tachyon deformation contributes as a 2d scalar potential on the worldsheet which acts as a barrier potential where light string states cannot propagate. A detailed study of general tachyon condensates goes well beyond string perturbation theory, so we only consider the first tachyonic deformation that arises when the asymptotic size of the circle of the BoN is taken to be small. This allows us to describe the early stages of tachyon condensation, employing techniques from \cite{Adams:2005rb}. In this picture, as the winding tachyon acquires a VEV, it spontaneously breaks the winding U(1) symmetry. In the dual picture, tachyonic KK modes condense, spontaneously breaking the momentum U(1) symmetry and making it so that the dual solution depends non trivially on the circle. The T-dual picture describes a configuration where the bulk tachyon locally condenses, ending the 26-dimensional spacetime. This points to the fact that the BoN might be another example where two cobordisms to nothing are T-dual to each other and that there is never a spontaneous decompactification.

Of course, this is only meant to be taken as a possible scenario, since we have ignored all of the other winding modes that come to deform the worldsheet in this limit. Nevertheless, this scenario is reminiscent of other examples in the literature where bubble of nothing and tachyon condensates can be seen as two sides of the same coin \cite{Fabinger:2000jd}. We therefore hope that this serves as motivation for a better understanding the full-fledged process of tachyon condensation: it seems to play a crucial role in providing physical mechanisms for cobordisms to nothing in theories without supersymmetry. In some cases, this can be tractable, such as in the $\alpha'$-exact examples in \cite{Hellerman:2006ff}.\\

Following an introduction to T-duality, we outline the Buscher rules in Section \ref{sec:tduality}. We then present the BoN in bosonic string theory in Section \ref{sec:bon} and apply the Buscher rules to it in Section \ref{sec:buscherules}. Subsequently, we contrast the result generated by the Buscher rules with what one would expect from the full T-duality. We argue that the Buscher rules provide a smeared approximation of the true solution. We find indications of this at the level of the gravity equations of motion, in section \ref{sec:gravityevidence}. Then, we describe the lower-dimensional perspective in section \ref{sec:dimred} after performing a dimensional reduction on the circle and we show that the dual solution has finite tension, indicating that it does not describe a spontaneous decompactification. Finally, in Section \ref{sec:wsevidence}, we turn to the worldsheet of the BoN and its T-dual. We explain how the T-dual solution develops a dependence on the dual circle due to winding modes becoming light. We then describe one scenario for a stringy description of the BoN, where the dual solution describes a cobordism to nothing of bosonic string theory through bulk tachyon condensation. We end this note with a few remarks on future directions and extensions to more phenomenologically relevant cases.

\section{Bosonic String Theory, T-duality and Buscher's Rules}\label{sec:tduality}

Bosonic string theory describes a string propagating in a 26-dimensional target space, whose massless degrees of freedom are simply given by a graviton, the dilaton and the Kalb-Ramond 2-form (the B-field). Bosonic string theory also contains a tachyon that has a string scale negative squared mass; it signals an instability of the 26-dimensional vacuum \cite{Polchinski:1998rq}. As is common practise when considering bosonic string theory, we will first assume in this note that the bulk tachyon is at the maximum of its potential, with a vanishing VEV. We can therefore describe bosonic string theory at low energies and weak coupling with the following action in the Einstein frame:\footnote{We consider this action to be valid on time scales shorter that the typical time scale for bulk tachyon condensation to take place.}
\begin{equation} \label{eq:actionbos}
    S= \frac{1}{2\kappa^2}\int d^{26} x \sqrt{-G} \{R - \frac{1}{6}(\partial \phi)^2 - \frac{1}{12} e^{-\phi/3} H_{\mu\nu\lambda}H^{\mu\nu\lambda}\}
\end{equation}
with gravitational coupling $\kappa^2 = 8 \pi G_N \sim l_s^{24} g_s^2$, where $G_N$ is the 26-dimensional Newton's constant, $l_s$ is the string length and $g_s$ is the string coupling, given by the constant part of the dilaton.

In order to discuss T-duality, we first briefly overview the compactification of the bosonic string on a circle. The resulting 25-dimensional theory has two U(1) gauge symmetries: one U(1) vector $A_\mu\sim G_{\mu\,25}$ arises from the the off-diagonal components of the 26-dimensional metric and another $\tilde{A}_\mu \sim B_{\mu\, 25}$ arises from the 26-dimensional Kalb-Ramond field with one leg extended in the 26-th dimension. The remainder of the components of the 26-dimensional metric and B-field yield a graviton and B-field in 25 dimensions, as well as an extra scalar, the radion $\sigma \sim G_{25\, 25}$. Finally the 26-dimensional dilaton, as a scalar, simply becomes a 25-dimensional dilaton after dimensional reduction. The field content of the resulting 25-dimensional theory is therefore given by a graviton, a B-field, a dilaton, a radion and two U(1) gauge fields. 

In the above discussion, we have assumed that all the 26-dimensional fields were independent of the 26-th coordinate that has been compactified. If we relax this constraint, we need to expand each field in Fourier modes around the circle:
\begin{equation}\label{eq:kkmode}
    \Phi (x^\mu; x^{25}) = \sum_{n=-\infty}^{\infty}\Phi_n(x^\mu) e^{i n x^{25}/R}\, ,
\end{equation}
where R is the radius of the circle and where this decomposition is meant to describe any field with all potential spacetime/gauge indices suppressed. The previous discussion only considered the zero-modes $\Phi_0$. The reason for this is that all of the higher order modes become massive: expanding the kinetic term of the field $\Phi$, one finds that the mode $\Phi_n$ acquires a mass proportional to $\frac{n}{R}$. These modes are KK modes and they can be ignored if we consider energies smaller than the KK scale given by $M_{KK}=R^{-1}$. One can easily show that they are charged under $A_\mu$ by performing a gauge transformation $x^{25}\to x^{25} + \Lambda (x^\mu)$ (see e.g. \cite{Tong:2009np}). Indeed, under such a transformation, the KK modes are shifted:
\begin{equation}\label{eq:shift}
    \Phi_n \to \Phi_n e^{i n \Lambda / R}\, , 
\end{equation}
which signals that the $n^{\text{th}}$ KK mode has charge $n/R$ under the momentum U(1). 

Taking a closer look at how the compact circle modifies the worldsheet theory, we notice that there is another infinite set of massive states: the winding modes. Indeed, since closed strings are extended objects, they not only have momentum modes along the circle but they can also display winding modes around the circle. One can show that these winding modes have a mass proportional to $M_W\sim (\alpha')^{-1} R$ and that they are charged under $\tilde{A}^\mu$. The massless fields in 25 dimensions are therefore both KK and winding zero-modes and as such are uncharged under the two U(1) symmetries. 

T-duality is an exact symmetry of the whole spectrum of the bosonic string compactified on a circle. The mass of a the string state with $n$ units of momentum along the circle and $m$ units of winding around the circle is given by \cite{Tong:2009np}:
\begin{equation}\label{eq:massformula}
    M^2 = \frac{n^2}{R^2} + \frac{m^2 R^2}{\alpha'^2}+ \frac{2}{\alpha'}( N + \tilde N -2) \,,
\end{equation} subjects to the so-called level-matching condition:
\begin{equation}
    N - \tilde N = nm\,.
\end{equation}
The first term tells you that the KK modes have a mass of $n\, R^{-1}$, and the second tells you that winding modes have a mass $m \,R\,(\alpha')^{-1}$. The massless states are obtained with zero momentum and zero winding, at level $N=\tilde N = 1$. They are: the metric, the B-field, the dilaton, the two U(1) vector fields and the radion. At zero momentum and zero winding, one also finds the bulk tachyon at level 0 that we have swept under the rug (for now). T-duality maps the spectrum of the bosonic string on a circle of radius $R$ to that of a bosonic string on a circle of radius $\alpha' R^{-1} $ by performing the following changes simulatneously: 
\begin{equation}\label{eq:Tduality}
    R \leftrightarrow \alpha' R^{-1} \,,\;\;\; n\leftrightarrow m\,.
\end{equation}
It is straightforward to check that the spectrum \eqref{eq:massformula} is invariant under this transformation. In this sense, bosonic string theory on a circle is  a self-dual theory under T-duality.

The winding modes for the original theory are therefore the KK modes of the dual. We can thus expand fields analogously to \eqref{eq:kkmode} to get the winding modes:
\begin{equation}\label{eq:windingmode}
    \Phi (x^\mu; \tilde{x}^{25}) = \sum_{n=-\infty}^{\infty}\Phi_m(x^\mu) e^{i m R\, \tilde{x}^{25}{\alpha'}^{-1}}\,.
\end{equation}
It is easy to see that as the $n^{\text{th}}$ KK modes has charge $n/R$ under the momentum U(1) \eqref{eq:shift}, the $m^{\text{th}}$ winding mode has charge $m \,R\, \alpha'^{-1}$ under the winding U(1). 

In practise, one usually only considers how T-duality acts on the zero-modes of the 26-dimensional fields. These zero modes correspond to the parts of the 26-dimensional fields that do not propagate at all in the compact dimension. The laws that describe how these fields transform under a T-duality are called the Buscher rules \cite{Buscher:1987sk}. We now describe them briefly. The Buscher rules for a circle compactification along a dimension parametrized by $\theta$ relates the B-field, metric and dilaton in the string frame of the original theory to those of the dual theory:

\begin{align}\label{eq:BuscherRules}
\begin{split}
    &\tilde{(G_s)}_{\theta\theta}= \frac{1}{(G_s)_{\theta\theta}}\,, \;\; \;\;\tilde{(G_s)}_{\theta\mu}= \frac{(B_s)_{\theta \mu}}{(G_s)_{\theta\theta}}\,, \;\;\;\; \tilde{(B_s)}_{\theta\mu}= \frac{(G_s)_{\theta \mu}}{(G_s)_{\theta\theta}}\,,\\  &\tilde{(G_s)}_{\mu\nu} = (G_s)_{\mu\nu} - \frac{(G_s)_{\theta \mu}(G_s)_{\theta \nu}-(B_s)_{\theta \mu}(B_s)_{\theta \nu}}{(G_s)_{\theta \theta}} \,,\\&\tilde{(B_s)}_{\mu\nu} = (B_s)_{\mu\nu} - \frac{(G_s)_{\theta \mu}(B_s)_{\theta \nu}-(B_s)_{\theta \mu}(G_s)_{\theta \nu}}{(G_s)_{\theta \theta}}\, , \\
    &\tilde \phi = \phi -\frac{1}{2}\log |(G_s)_{\theta\theta}|\,,
    \end{split}
\end{align}
where the coordinates labeled $\{\mu,\nu\}$ correspond to the transverse coordinates to the circle and where the $s$ subscripts are meant to emphasize that these fields are evaluated in the string frame. This is the frame where the Einstein-Hilbert term is multiplied by a factor of $e^{- 2 \phi}$ which reflects the fact that the action has been computed at tree-level in string perturbation theory. It is convenient to work in the more familiar Einstein frame, where the Einstein-Hilbert term is not multiplied by the dilaton. One can obtain this by performing a conformal re-scaling of the string frame metric, which in 26 dimensions is given by: 
\begin{equation}\label{eq:framechange}
    G_s= e^{\frac{\phi}{6}}G\,,\;\;\;  \tilde{(G_s)}= e^{ \frac{ \tilde \phi}{6}}\tilde G\,.
\end{equation}
Equipped with the Buscher rules we are thus ready to T-dualize any solution to the action \eqref{eq:actionbos} along a circle parametrized by $\theta$. In the next sections we will proceed to applying them to the BoN in 26 dimensions. With foresight, it is important to remember that the Buscher rules are blind to all of the physics surrounding the KK and winding modes; they are only a good proxy for describing T-duality when all of these modes are heavy and decouple from the low-energy physics.

\section{The Bubble of Nothing}\label{sec:bon}

We now turn to the subject of interest: the bubble of nothing (BoN). The original BoN \cite{Witten:1981gj} was built in five dimension but it is easily generalized to any number of spacetime dimensions, and in particular to 26 of them. We now describe this geometry before considering how it behaves under T-duality in section \ref{sec:TdualBoN}. 

The BoN is a solution of Einstein's equations in the vacuum and as such it is a solution of the action \eqref{eq:actionbos} if we take the dilaton to be constant and turn off the B-field. The metric of the BoN in 26 dimensions is given by: 
\begin{equation}\label{eq:BoNsol}
    ds^2 = f(r)^{-1} dr^2 + (g_{bon})_{ij} dx^i dx^j + R_0^2 f(r) d\theta^2 \, ,
\end{equation}
with $\theta \in \{0,2\pi\}$ and with the metric on the $24$-dimensional bubble wall is given by: 
\begin{equation}
    (g_{bon})_{ij} dx^i dx^j = r^2 \{- dt^2 + \cosh(t)^2 d\Omega^{2}_{23}\}\, .
\end{equation}
The warp factor $f(r)$ is defined by: 
\begin{equation}\label{eq:f23}
    f(r)= 1- \left(\frac{r_0}{r}\right)^{23}\,.
\end{equation}
This bubble has the exact same features as the original 5-dimensional one \cite{Witten:1981gj}. Firstly, by rotating to euclidean time $t \to i \varphi + \pi/2$, we get the euclidean metric:
\begin{equation}\label{eq:BoNeucl}
    ds^2 = f(r)^{-1} dr^2 + r^2 ( d\varphi^2 + \sin(\varphi)^2 d\Omega_{23}^2)+ R_0^2 f(r) d\theta^2 \,,
\end{equation}
from which it is clear that when $r\to \infty$ the metric describes flat euclidean space compactified on a circle of radius $R_0$:
\begin{equation}\label{eq:BoNatinf}
    ds^2 = dr^2 + r^2 d\varphi^2 + r^2 \sin(\varphi)^2 d\Omega_{23}^2+ R_0^2  d\theta^2 \,.
\end{equation}
Furthermore, we see from \eqref{eq:f23} that the circle parametrized by $\theta$ pinches off at $r=r_0$, effectively ending spacetime. Importantly, this point is non-singular for a specific choice of $R_0$ (or equivalently a specific choice of $r_0$). Indeed, by expanding the metric around $r=r_0$ and by making the change of variable $\lambda^2 = r - r_0$, we see that the metric in the $\{r,\theta\}$-plane reduces to: 
\begin{equation}
    \frac{4 r_0}{23} \left( d\lambda^2 + \left(\frac{ 23\, R_0 }{2 \,r_0}\right)^2 \lambda^2 d\theta^2\right)\,.
\end{equation}
This describes the smooth 2-dimensional plane in polar coordinates if and only if $R_0$ depends on $r_0$ as:\footnote{For a general dimension d, one finds $R_0= 2 r_0 (d-3)^{-1}$, which explains why $R_0=r_0$ in the original 5-dimensional BoN.}
\begin{equation}\label{eq:R0r0}
    R_0 = \frac{2}{23} r_0\,.
\end{equation}
This can also be seen from evaluating curvature invariants for the metric \eqref{eq:BoNatinf}. Being a solution in the vacuum, the Ricci scalar vanishes everywhere. We therefore turn to the Kretschmann scalar which gives: 
\begin{equation}
    K = R_{\mu\nu\delta\gamma}R^{\mu\nu\delta\gamma}\propto  \frac{r_0^{46}}{r^{50}}\,,
\end{equation}
which is finite at $r=r_0$, signalling that there is no curvature singularity there. 

This means that the geometry is smooth at $r=r_0$ and curvature corrections can be ignored everywhere. Importantly, since the dilaton is constant, the size of the circle goes to zero in string units as $r\to r_0$. As a result, if $R_0$ is large in string units and the string coupling is small, the BoN describes a smooth cobordism to nothing, entirely described within the realm of validity of the low-energy EFT of bosonic string theory. Stringy physics will come into play if the size of the circle $R_0$ becomes comparable to the string scale. This fact will play an important role in section \ref{sec:wsevidence} where we study the T-dual of the BoN from the worldsheet perspective.   

\section{The T-dual Solution}\label{sec:TdualBoN}

We now go to the heart of the subject and consider the T-dual of the BoN. Before applying Buscher's rules to the BoN \eqref{eq:BoNsol}, let us first comment on what one should expect from T-dualizing the BoN along the circle that pinches off.

As detailed in the previous section, the BoN describes a circle compactification where the circle smoothly pinches off on the surface of the bubble wall. The metric \eqref{eq:BoNsol} enjoys a U(1) isometry which acts by shifting the value of $\theta$ and ensures that momentum around the circle is conserved. However, string winding number cannot be conserved since the string can slip off by going to the point of pinch-off, as shown schematically in figure \ref{figure1}. Since T-duality exchanges momentum and winding states we thus expect the T-dual solution to have a U(1) winding symmetry but no conserved momentum U(1) along the circle. This would mean that the solution depends non-trivially on the angle $\theta$ parametrizing the dual circle. 

In the next section we apply Buscher's rules \eqref{eq:BuscherRules} to the BoN solution \eqref{eq:BoNsol}. We will find that the resulting solution displays a momentum U(1) isometry and thus contradicts our expectations. We will then detail how this apparent puzzle is solved in section \ref{sec:clues}, before turning to the worldsheet perspective in section \ref{sec:wsevidence}.

\subsection{Naively applying Buscher's rules}\label{sec:buscherules}
Applying the Buscher rules to the BoN is done in three steps: first one puts \eqref{eq:BoNsol} in the string frame, then one applies the rules \eqref{eq:BuscherRules} and finally one must go back to the Einstein frame in the dual theory. The solution is given by the following metric and dilaton: 
\begin{equation}\label{eq:dualbon}\begin{gathered}
    d\tilde s ^2 = \left(e^{\frac{\phi_0}{6} }R_0^2 f(r)\right)^{\frac{1}{12}} \left[f(r)^{-1} dr^2 + (g_{bon})_{ij} dx^i dx^j \right] + e^{-\frac{23}{72}\phi_0}\left(R_0^2 f(r)\right)^{-\frac{11}{12}} d\tilde{\theta}^2  \\
    \tilde \phi = \frac{11}{12}\phi_0 -\frac12\log\left[f(r) R_0^2\right] \,.
\end{gathered}\end{equation}
This solution is very different from the original BoN. Firstly, it features a running dilaton which blows up at $r=r_0$ and takes us to strong coupling. On top of that, there is a curvature singularity at $r=r_0$, as can be seen by computing the Ricci scalar: 
\begin{equation}\label{eq:curvaturescalar}
G_{\mu\nu}R^{\mu\nu} \sim (R_0^2 f(r))^{-\frac{13}{12}}  \left(\frac{r_0}{r}\right)^{48}\,.
\end{equation}
It is immediate to see that at $r=r_0$, this quantity blows up. From now on, for simplicity, we take $R_0$ to be a dimensionless quantity, and we grant a dimension of length to the coordinate $\theta$. As the intuition regarding T-duality would indicate, this solution is one where the circle becomes infinite-sized in string units as $r\to r_0$:\begin{equation}\label{eq:dualcirclesize}
    \frac{R^2(S^1)}{l^2_s}\sim \left(R_0^2 f(r)\right)^{-\frac{11}{12}} e^{\frac{\tilde \phi}{ 6}}  \sim (R_0^2 f(r))^{-1}  \xrightarrow[r\to r_0]{} \infty\,.
\end{equation} 
At spatial infinity, one can see that the circle asymptotes to the expected value in string units: \begin{equation} \label{eq:Asymptradidual}
    \frac{R(S^1)}{l_s}\xrightarrow[r\to \infty]{} R_0^{-1} e^{-\frac{\phi_0}{ 12}} \,.
\end{equation} 
Note that in the original frame $R(S^1) \,l_s^{-1}\xrightarrow[]{r\to\infty} R_0 \,e^{\frac{\phi_0}{12}}$ in string units, so the quantity \eqref{eq:Asymptradidual} is exactly what one should expect.

Summing up, the dual solution given by the Buscher rules describes what one would naively expect from T-duality: the circle blows up instead of pinching off. However, this solution is strongly coupled and singular at the core $r=r_0$. As such, it is expected to be heavily corrected as soon as one gets too close to $r=r_0$. Furthermore, as advertised, this solution does not depend on the angular coordinate and thus seems to display a momentum U(1). This contradicts our expectations from T-duality exchanging winding and momentum states, given that the BoN does not display a winding U(1). All of these elements point to the fact that the solution \eqref{eq:dualbon} is not the full picture. Before going to the worldsheet perspective in section \ref{sec:wsevidence}, we will give further clues at the level of gravity that there must be more to the story. 

\subsection{Clues from spacetime }\label{sec:clues}
There are other instances in the literature where the Buscher rules fail to provide an accurate picture of the dual solution. For instance, the case at hand is very similar to that of the T-duality between the NS5 brane and the Taub-NUT geometry discussed in \cite{Gregory:1997te,Tong:2002rq}. The Taub-NUT geometry is akin to that of the BoN in that it also describes a circle smoothly pinching off in spacetime. Like the BoN, it features a momentum U(1) isometry but string winding number around the circle is not conserved since the string may slip off, either by moving to the point where the circle degenerates, or alternatively by wrapping once around the asymptotic boundary of Taub-NUT. The Taub-NUT geometry is known to be T-dual to a single NS5-brane \cite{Callan:1991dj} localized on the dual circle. At first, this relationship was only proven for a \textit{smeared} NS5-brane, for which the transverse circle is an isometry and in which case the Buscher rules can be applied. Indeed the metric obtained from T-dualizing the Taub-NUT geometry with the Buscher rules, much like \eqref{eq:dualbon}, enjoys a U(1) isometry around the circle. 

In \cite{Tong:2002rq}, the T-duality between the Taub-NUT geometry and a \textit{localized} NS5 brane was derived. This is done by studying the worldsheet theory that corresponds to the Taub-NUT geometry. They find that although the naive T-duality transformation yields a metric with a U(1) momentum isometry, there are worldsheet instanton contributions that break this isometry such that the solution depends on $\theta$. This can be mapped exactly to the supergravity solution of an infinite, periodic, array of colinear NS5 branes, which mimics a localized NS5 brane on a compact circle. 

It is natural to imagine that something similar happens for the T-dual solution of the BoN. It could be that the true solution depends non trivially on the dual circle. The Buscher rules would then only provide the smeared version of this solution. 

In the remainder of this section, we will give more quantitative evidence that the solution \eqref{eq:dualbon} is but an approximation of the true solution. In section \ref{sec:gravityevidence}, we will see that the equations of motion indicate that the metric \eqref{eq:dualbon} describes an object that is smeared along the angular dimension. Then, in section \ref{sec:dimred}, we will argue that the dual solution has finite tension, indicating that it cannot describe a spontaneous decompactification. 

\subsubsection{A smeared solution in gravity} \label{sec:gravityevidence}
We now describe evidence to support the fact that the solution obtained in \eqref{eq:dualbon} is smeared along the angular coordinate. We can do this by interpreting the geometry described by \eqref{eq:dualbon} as describing the back-reaction of a source term and showing that the structure of the equations is such that the source term has to be localized in $r$ and span $\tilde{\theta}$.

Indeed, although the solution is singular and we do not know its UV resolution, we can attempt to describe it as a localized source term in the action \eqref{eq:actionbos} as: 
\begin{equation}\label{eq:surfacedelta}
    -\lambda \int d^{26}x \sqrt{- g_{sm}} \, e^{a \tilde \phi} \, \delta(\vec{x})\, ,
\end{equation}
where $g_{sm}$ is the pull-back of the ambient metric on the world-volume of this source, $\lambda = 2 \kappa^2 T$ is a tension and where we have introduced a Dirac delta function to schematically describe the point-like nature of the source (but we will not be careful about its precise definition). We have introduced a general exponential coupling to the dilaton. The equations of motion with the source term can be obtained from \eqref{eq:actionbos} and lead to: 
\begin{equation}\begin{gathered}
    R_{\mu\nu} - \frac{1}{2} G_{\mu\nu} + \frac{1}{12} G_{\mu\nu} (\partial  \tilde \phi)^2 - \frac{1}{6} \partial _ \mu \tilde \phi \partial _ \nu \tilde \phi = - \lambda  \delta^i _ \mu \delta ^j _ \nu  (g_{sm})_{ij}\sqrt{\frac{g_{sm}}{G}} e^{ a \tilde \phi}\delta(\vec{x})\, ,\\
    \frac{1}{3}\partial _ i \left(\sqrt{-g_{sm}} (g_{sm})^{ij} \partial _ j \tilde \phi \right) = 2 a \lambda \sqrt{-g_{sm}}e^{a \tilde \phi} \delta(\vec{x})\, .
\end{gathered}\end{equation}
where the $\{\mu,\nu\}$ indices span all of the 26 dimensions. We see that the delta function only appears in the dilaton equation of motion (if $a\neq 0$) and in the components of the Einstein equations along which the object is \textit{extended}.

Since we do not yet know along which directions this object might be (or not) localized, we study the equations of motion for this dual solution in the vacuum before discussing potential source terms. Consider the ansatz \eqref{eq:dualbon}, keeping $f(r)$ general. The equations of motion in the vacuum are given by:
\begin{equation}
    \begin{aligned}
       \{r,r\}:\;\;\;\; &-23 + 23 f(r) + r f'(r) =0\,,\\
       \{\theta,\theta\}:\;\;\;\; &24 f'(r)+r f''(r) =0\,,\\ \{i,j\}:\;\;\;\; & 24 f'(r)+r f''(r)=0\,,\\
       \tilde \phi:\;\;\;\; &24 f'(r)+r f''(r)=0\,.
    \end{aligned}
\end{equation}
If we want to introduce a source term on the right-hand side of the equations, we have to make sure that the structure on the left-hand side is capable of generating a delta function. Delta functions arise when there is a discontinuity in the derivatives of the metric components, as per the Israel Junction conditions \cite{Israel:1966rt}. Indeed, this discontinuity of the first derivatives leads to a delta function for the second derivatives. The first equation only has first derivatives, so it is safe to assume that it does not source a delta function. We infer from this that the object has to be localized in the radial direction, which fits with our understanding of it.

The three others, however, are exactly the same and contain second derivatives. This means for instance, that the dilaton is sourced by this object ($a\neq 0$). This should not come as a surprise since it has a non-trivial profile in \eqref{eq:dualbon}. More interestingly, it would seem that there is a source term along the angular coordinate which indicates that the solution is extended along $\tilde \theta$. This fits perfectly with the picture that the solution \eqref{eq:dualbon} is describing an object that is smeared along the angular direction. 

We could now try to solve the equations explicitly, with a source term, where the metric $g_{sm}$ is given by the ambient metric \eqref{eq:dualbon} for all coordinates except the radial one. This would allow us to determine $\lambda$, the tension of the smeared solution, as well as $a$, its dependence on the dilaton. One can try to solve the equations but this fails: the choice $a=\frac{13}{12}$ allows one to match the divergences on both sides but we find 3 different values of $\lambda$ for each of the three equations. We interpret this as the fact that the smeared solution cannot be described by an ansatz as simple as \eqref{eq:surfacedelta} where only a source term for the dilaton is introduced; a more complicated source term may be a better fit.

In order to determine the tension of the object, a key insight is that once compactified on the circle, the angular dependence of the metric should wash out completely. The smeared solution and the localized one should therefore look the same from the lower-dimensional perspective. In the next section \ref{sec:dimred}, we use this fact to derive a \textit{finite} tension for the dual object in 25 dimensions. 

\subsubsection{Finite tension from dimensional reduction}\label{sec:dimred}
From the perspective of the theory compactified on the circle, we can ignore the dependence of the solution on $\theta$, and we can describe the dual object as a localized tension at $r=r_0$. Indeed, this means that from the lower-dimensional perspective, the smeared solution \eqref{eq:dualbon} is a good proxy for describing the dual object, since all angular dependence drops out anyways due to the dimensional reduction. We therefore proceed with the dimensional reduction of the smeared solution. We take the following compactification ansatz:
\begin{equation}\label{eq:compactansatz}\begin{aligned}
    ds^2 &= e^{- 2 \alpha \omega(r) } ds^2_{25} + e^{2 \beta \omega(r)} d\theta ^2\, ,\\
    \phi &= \phi(r)\,.
\end{aligned}\end{equation}
By requiring the Einstein frame and a canonically normalized kinetic term for the radion in 25 dimensions, we have: 
\begin{equation}
    \gamma = \frac{\alpha}{\beta} = \frac{1}{23}\,,\;\; \beta^2= \frac{23}{24}\,.
\end{equation}
The lower-dimensional theory has the following action, where we have turned off all fields irrelevant for our discussion:
\begin{equation}\label{eq:action25}
    S_{25}= \frac{1}{2\kappa_{25}^2} \int d^{25}x \sqrt{-G_{25}} \{R_{25} - \frac{1}{6} (\partial \tilde \phi)^2 - (\partial \omega)^2\}
 - T \int d^{25} x \sqrt{-g_{24}} e^{ a \tilde \phi}  e^{ b \omega}  \;\delta(r-r_0)\end{equation}
where we have already put in a source term in preparation for what comes next. Note that $\lambda = 2 \kappa_{25}^2 T$ has dimension of energy. To obtain the dimensionally reduced solution, we can match the ansatz \eqref{eq:compactansatz} to the metric \eqref{eq:dualbon} to find: 
\begin{align}
    e^{2 \beta \omega(r)}  &=  e^{-\frac{23}{72}\phi_0}\left(R_0^2 f(r)\right)^{-\frac{11}{12}}  \label{eq:solradion}\\
    ds^2_{25} &= e^{2 \alpha \omega(r)}\left(  \left(e^{\frac{\phi_0}{6} }R_0^2 f(r)\right)^{\frac{1}{12}} \left[f(r)^{-1} dr^2 + (g_{bon})_{ij} dx^i dx^j \right]\right) \label{eq:solmetricomp}\\
    \text{ }&= \left(R_0^2 f(r)\right)^{\frac{1}{23}} \left[f(r)^{-1} dr^2 + (g_{bon})_{ij} dx^i dx^j \right]\label{eq:dualboncompact}\\
    \tilde \phi &= \frac{11}{12}\phi_0 -\frac12\log\left[f(r) R_0^2\right]
\end{align}
This should be a solution to the 25-dimensional equations of motion with a localized source term at $r=r_0$. For a special choice of $a$ and $b$, the equations can be solved simultaneously as: 
\begin{equation} \label{eq:lowerdimdual}
    \begin{aligned}
       \{r,r\}:\;\;\;\; &-23 + 23 f(r) + r f'(r) =0\,,\\
       \{i,j\}:\;\;\;\; &24 f'(r)+r f''(r) = - 2 \lambda \frac{r_0}{R_0}\delta(r-r_0)\,,\\ \omega:\;\;\;\; & 24 f'(r)+r f''(r) = - 2 \lambda \frac{r_0}{R_0}\delta(r-r_0)\,,\\
       \tilde \phi:\;\;\;\; &24 f'(r)+r f''(r) = - 2 \lambda \frac{r_0}{R_0}\delta(r-r_0)\,.
    \end{aligned}
\end{equation}
The very special choice of a and b that allows for this drastic simplification is given by: 
\begin{equation}
    a = \frac{1}{6} ,\;\; \text{and} \; b = \frac{11}{\sqrt{138}}\,.
\end{equation}
This odd value of b can be partially explained by the fact that the kinetic term  of the radion is normalized to one. The first equation in the vacuum can be solved to yield the correct profile for $f(r)$ \eqref{eq:f23}. The last three equations in \eqref{eq:lowerdimdual} can be solved by integrating:
\begin{equation}
   \int dr\{ 24 f'(r)+r f''(r)\}= - 2 \lambda \frac{r_0}{R_0} 
\end{equation}
This can be solved to give: 
\begin{equation}
    f(r) = -\frac{2  r_0}{23 R_0} \lambda+ \frac{c_0}{r^{23}} = - \lambda + \frac{c_0}{r^{23}}
\end{equation}
where we have used \eqref{eq:R0r0} in the second equation. Therefore, for the consistency of the solution, we must choose the integration constant to be $c_0 = -r_0^{23}$ and $\lambda = - 1$. The result of this computation is that from the lower-dimensional perspective, the 24-dimensional object has a finite tension of 
\begin{equation}\label{eq:tension}
    \lambda = -1 \;\;\to\;\;T = -(2 \kappa^2_{25})^{-\frac{24}{23}} \, ,
\end{equation}
and couples to the dilaton with a coupling $a= \frac16$.

One can perform a similar analysis for the BoN in the original frame. Indeed, the two solutions should be identical from the lower-dimensional perspective. The action is the same as above in \eqref{eq:action25}, except that the dilaton is $\phi = \phi_0$ and that $a=0$ since the BoN does not source the dilaton. For the compactified BoN, we dub the radion $\nu$ so as to avoid confusion and we take the source term: \begin{equation}
    - T \int d^{25} x \sqrt{-g_{24}}e^{ c \nu}  \;\delta(r-r_0)\,.
\end{equation}
The solution is obtained analogously to the previous example. One obtains:
\begin{align}
    e^{2 \beta \nu(r)}  &= R_0^2 f(r) \, , \\
    ds^2_{25} &= e^{2 \alpha \nu(r)}\left[f(r)^{-1} dr^2 + (g_{bon})_{ij} dx^i dx^j \right] \, ,\\
    \;\text{ } &= \left(R_0^2 f(r) \right)^{\frac{1}{23}} \left[f(r)^{-1} dr^2 + (g_{bon})_{ij} dx^i dx^j \right]\, , \label{eq:boncompact}\\
     \phi &= \phi_0\, , 
\end{align}
with the same values of $\alpha$ and $\beta$ since the dimensional reduction is exactly the same. Solving the equations of motion simultaneously is only possible if $c= - 2 \sqrt{\frac{6}{23}}$: 
\begin{equation} \label{eq:lowerdimdualbon}
    \begin{aligned}
       \{r,r\}:\;\;\;\; &-23 + 23 f(r) + r f'(r) =0\,,\\
       \{i,j\}:\;\;\;\; &24 f'(r)+r f''(r) = - 2 \lambda \frac{r_0}{R_0}\delta(r-r_0)\,,\\ \omega:\;\;\;\; & 24 f'(r)+r f''(r) = - 2 \lambda \frac{r_0}{R_0}\delta(r-r_0)\,.
    \end{aligned}
\end{equation}
We fall back on the same equations of motion as for the dual solution. In particular, the two solutions have the same tension \eqref{eq:tension}. Furthermore, we see from \eqref{eq:boncompact} and \eqref{eq:dualboncompact} that the two metrics are exactly the same. In order to truly match the two solutions, the only thing missing is to match the scalars. In the case of the BoN, the radion is the only field that runs whilst in the dual frame, both the dilaton and the radion run. This is solved by showing that the radion in the BoN frame is a linear combination of the dilaton and radion in the dual frame: 
\begin{equation}
    c\; \nu(r) = a\; \tilde \phi(r) + b\; \omega(r)\,.
\end{equation}
The two solutions therefore match perfectly in 25 dimensions, as should be expected from T-duality. On top of being a nice consistency check, this analysis gives us further evidence that the 24-dimensional object that is T-dual to the BoN has a finite tension. In particular, once compactified on the circle, we have shown that it has a tension of $O(1)$ in 25-dimensional Planck units. 

Note that even if the BoN described a perfectly smooth geometry in 26 dimensions, it is singular once compactified on the circle. Indeed, one can check that the point $r=r_0$ is the locus of a curvature singularity for the metric \eqref{eq:boncompact}. This was already noted in \cite{Anous:2019rqb} where the BoN was described as dynamical cobordism to nothing. The dual solution on the other hand is singular both in 25 and in 26 dimensions, as well as strongly coupled.\\

This ends our discussion of the gravity description; we have given quantitative evidence that the solution \eqref{eq:dualbon} is too naive to describe the true T-dual of the BoN. We now go beyond this low-energy discussion and discuss the T-duality of the BoN from the perspective of the worldsheet.

 \subsection{The worldsheet perspective}\label{sec:wsevidence}
By implementing the Buscher rules in section \ref{sec:buscherules}, we have only considered how the zero-modes of the 26-dimensional fields under dimensional reduction transform under T-duality. We have seen that this yields a limited, approximated description of the T-dual object. Indeed, T-duality is a duality of the whole spectrum of the bosonic string on a circle, as emphasized in \eqref{eq:Tduality}. It must therefore be that some of the KK and winding modes are playing a central role in this T-duality transformation. For this reason, we now take the perspective of the worldsheet, where we can try to keep track of the whole spectrum of the string.  

We are interested in the geometry of the T-dual solution, we therefore want to make the dual circle large in string units. This corresponds to considering the original BoN with an asymptotic radius $R_0$ that is substringy, as per \eqref{eq:Tduality}. This means that we cannot consider the simple BoN solution in Eq. \eqref{eq:BoNsol} but instead, we have to consider a stringy version of the BoN. We now take a step in this direction, by considering the BoN worldsheet sigma model with large radius $R_0$ at $r\to \infty$ and studying how this theory is deformed when $R_0$ becomes string-sized.

The explicit worldsheet theory corresponding to the BoN, has, as far as we know, not been explicitly constructed in the literature. Nevertheless, since the BoN is weakly coupled everywhere and non-singular, we expect to be able to describe it using standard perturbative methods. We will not endeavor to describe the exact world-sheet theory of the BoN, instead, we will consider the sigma model that describes the circle compactification part of the BoN, far away from the point of pinch-off. When the circle $R_0$ is large, general relativity applies and the BoN is described by a worldsheet sigma model with action in conformal gauge: 
\begin{equation}
    S= \frac{1}{4 \pi \alpha'}\int d^2 \sigma (G_s)_{\mu\nu} \,\partial_{a}X^{\mu} \partial ^a X^\nu + \text{ghosts}\,.
\end{equation}
Indeed, in this limit, the topology of the BoN is simply $\mathbb{R}^{25}\times S^1_R$. When one takes the limit where $R_0\to 1$, the size of the circle $R_0$ becomes stringy. In order to understand what is happening to the worldsheet theory in this limit, we can look at the mass spectrum of the bosonic string on a circle \eqref{eq:massformula}. We see that as $R_0$ becomes small, momentum modes become increasingly massive whilst winding modes become light. This signals the break-down of the effective description in target space, since all of the infinite tower of winding states are becoming massless. As expected from \eqref{eq:windingmode}, these states are going to depend on the coordinate parametrizing the dual circle. If any of these new modes develops a VEV, then the sigma-model will depend non-trivially on the dual circle: this would explain why the dual solution has this type of dependence. 

To be more explicit, we can consider a winding mode for the tachyon (though any other winding mode will also do the trick). We parametrize the worldsheet scalar corresponding to the radial direction by $X_r$, that corresponding to target space time by $X_t$ and that corresponding to the angular direction by $X_\theta$. The worldsheet scalar corresponding to the angular direction in the \textit{dual} theory is $\tilde X _{\theta}$. The winding tachyon vertex operator is simply given by:
\begin{equation}\label{eq:tachyon}
    T(X)=   T(r) e^{\kappa X_t}  e^{ i m \frac{R}{\sqrt{\alpha'}}\tilde X_{\theta}}  \,,
\end{equation}
where we have assumed that the tachyon carries momentum in time and has winding number $m$. The dependence on $X_r$ and $\tilde X_{\theta}$ corresponds to a generic winding mode of winding number $m$ \eqref{eq:windingmode}. Note that \textit{a priori}, since $R_0$ is small everywhere, there is no immediate reason why the tachyon should depend on $r$ in $T(r)$. However, since we expect the worldsheet description to have the same symmetries as that of the BoN, which depends on $r$, we will assume that this is the case.

The VEV \eqref{eq:tachyon} breaks the winding U(1) explicitly. Indeed, if one considers a gauge transformation $\tilde \theta \to \tilde \theta + \phi(x^{\mu})$, the tachyon \eqref{eq:tachyon} transforms as: 
\begin{equation}
    T(X) \to T(X) e^{i m \frac{R}{\sqrt{\alpha'}}\phi(x^\mu)}\,.
\end{equation}
In the geometric picture of the BoN, the winding U(1) was approximately exact at large $r$ far from $r=r_0$. However, once we make the circle small, the breaking of this symmetry is made explicit by the appearance of the winding tachyon. In this picture, $\phi(x^\mu)$ can be understood as a sort of Goldstone boson for the broken U(1) symmetry. As a side note, we remark that in the geometric description of the BoN, the U(1) is broken geometrically. To see this, recall that the winding U(1) gauge field $\tilde A ^\mu$ is obtained by dimensionally reducing the B-field and putting one of its ``legs'' along the circle. This comes down to decomposing the B-field as $B_2 = \tilde A \wedge dx$, where $dx$ is a one-form. For the BoN, $dx$ is not a harmonic form everywhere, it is only approximately harmonic at $r\to \infty$. In the asymptotic region, the $\tilde A ^\mu$ gauge field is therefore massless and the winding U(1) is approximately conserved. Close to the pinch-off point, $dx \neq 0$ and therefore the gauge field gets massive and the symmetry is broken.\\

We have thus seen how winding modes (tachyonic or not) are responsible for breaking the winding U(1) and generating a dependence on the dual circle. A less qualitative understanding of the dual of the BoN needs a full-fledged sigma-model description. In the geometric picture, when $R_0$ is large and the BoN is fully described by Einstein gravity: it is smooth and weakly coupled everywhere. This indicates that one should be able to make sense of the sigma-model describing it. The main difficulty that arises in doing so is the same as one encounters when attempting to quantize the string on any time-dependent background. It could be that there is a better choice of spacetime coordinates for which the BoN is static. Alternatively, one could perhaps engineer a static version of the BoN, by adding in extra ingredients so as to stabilize it. In both of these cases, identifying the correct CFT could be more tractable, and could perhaps be done using techniques akin to those of \cite{Papadopoulos:2002bg} and references therein. We leave this for future work.\\

As a last observation, we comment on one possible realization of a stringy version of the BoN. It could be that it is described by the condensation of the winding tachyon \eqref{eq:tachyon}. Indeed, condensing this tachyon can cause the system to lose the region where the circle is smaller than the string length, capping off the remaining region, in a stringy realization of the BoN. 

Let us first motivate this scenario by noting that the geometric bubbles of nothing and closed string tachyon condensation can often be seen as two sides of the same coin. A first example is provided by the FZZ duality \cite{FZZ} (see review in \cite{Kazakov:2000pm}) mentioned above. This duality is between the cigar CFT \cite{Witten:1991yr} and a Sine-Liouville theory with a potential. This potential on the worldsheet acts as a barrier beyond which the string modes cannot propagate. As such it provides a stringy version of the end of spacetime. The duality states that the Sine-Liouville theory is the better description (is weakly coupled) when the cigar is of sub-stringy size (and thus strongly coupled). Importantly, the Sine-Liouville potential can be seen as an expectation value of the closed string tachyon with winding one around the cigar's circle. This thus provides an example where a geometric pinch-off is replaced by a closed string tachyon condensate at the string scale.

Another famous example is that of $E_8\times \bar{E}_8$ theory in \cite{Fabinger:2000jd}, where the authors consider a compactification of M-theory on an interval where the two ends of the interval preserve different supersymmetries. Since supersymmetry is completely broken, the configuration is unstable and there is an attractive force between the two boundaries. This can be thought of as a closed-string version of the brane-anti-brane system that is known to develop an open-string tachyon instability and annihilate \cite{Sen:2004nf}. The authors conjecture that a similar thing happens to the $E_8\times \bar{E}_8$: that when the interval becomes of a size comparable to the eleven-dimensional Planck length, the system develops a tachyonic instability. On the other hand, when the interval is very large such that the Casimir force driving the two boundaries together can be treated as a perturbation, the system is shown to develop a bubble of nothing instability. This motivates how tachyon condensation can be seen as the high-energy analog of the bubble of nothing.

Furthermore, it is worth mentioning that this is not the first time that condensing winding tachyons has been used to describe topology-changing transitions in string theory. For instance, they have been considered in \cite{Adams:2005rb}, where a Scherk-Schwarz circle pinches off in spacetime. This describes a version of the BoN that is topologically allowed in theories with fermions since the boundary conditions on the circle are anti-periodic. This points to the fact that our analysis should extend to more general bubbles of nothing, in more realistic setups, with fermions. Similarly, in \cite{Adams:2001sv}, winding tachyons were used to smooth out singularities in non-compact non-supersymmetric orbifolds. In these situations the pinch-off point is singular from the perspective of general relativity but it gets capped off smoothly by a winding tachyon condensate. This opens the possibility that the considerations in this note might extend to singular dynamical cobordisms to nothing. \\

We now discuss in more detail how this possible scenario could be implemented. As noted above, when $R_0$ becomes of order of the string length, tachyonic winding modes such as \eqref{eq:tachyon} can come to deform the worldsheet theory. The situation at hand closely resembles that of \cite{Adams:2005rb}, albeit for a circle with periodic boundary conditions. We therefore borrow the tools and analysis developed there and use them for our purposes. We now show how condensing a winding tachyon can provide the stringy physics that smoothly cap-off the circle in the BoN, providing a stringy analog of the geometric BoN. Classically condensing the tachyon in real time comes down to adding an on-shell marginal tachyon vertex operator to the worldsheet action and solving the resulting path integral in the deep IR (see e.g. \cite{Hellerman:2006nx}):
\begin{equation}\label{eq:scpotential}
    S_T = \int d^2 \sigma \,  T(X)\,.
\end{equation}
We now assume that the tachyon gets a VEV as in \eqref{eq:tachyon}. By letting its VEV depend on $r$, $T(r)$, we see that the tachyon condensation process would start at some point in $r$ before consuming all of spacetime. This describes a stringy version of what happens in the geometric picture when the bubble of nothing nucleates at some finite point in $r$ at starts expanding.

We can now show that the presence of this tachyon creates a barrier potential on the worldsheet, keeping modes of the string from penetrating into the region of $r$ where the tachyon is condensed. The tachyon \eqref{eq:tachyon} couples to the bosonic string worldsheet as a scalar potential \eqref{eq:scpotential}: 
\begin{equation}\label{eq:potential}
    V= e^{\kappa X_t} \left(T(r) e^{ i m \frac{R}{\sqrt{\alpha'}}\tilde X_{\theta}} + \bar{T}(r)e^{- i m \frac{R}{\sqrt{\alpha'}}\tilde X_{\theta}}\right)\,,
\end{equation} where it is now explicit that the tachyonic winding mode $T(r)$ is actually always complex, since you can add winding and anti-winding modes with different phases. In the presence of the tachyon, the worldsheet becomes a non-trivial sigma model on a time-dependent target space: this is the tachyon condensate phase. Without going into the detail of solving this sigma model and studying the end-point of the tachyon condensation process, we get a qualitative understanding of how this potential affects the worldsheet. We see that, at late times, this potential exhibits a barrier to penetration into the region where the tachyon vertex operator has support, where $T(r) \neq 0$, such that we achieve a classical barrier for all values of $\tilde \theta$. From the spacetime perspective it means that spacetime ends ``smoothly'' at finite r, beyond which the massless string states become massive and there is no notion of space and time. 

A more thorough analysis would come down to taking quantum effects into account to see how the classical potential barrier is modified. This was done in \cite{Adams:2005rb} in a limit where the worldsheet theory resembles a supersymmetric sine-Gordon theory, for which the RG flow is known. No matter what, since we only took into account the first tachyonic deformation \eqref{eq:tachyon} and have not discussed any of other winding modes that deform the worldsheet theory in this limit, the potential \eqref{eq:potential} is expected to be heavily corrected and we cannot be sure that it paints the correct picture.

Finally, it is worth noting that although we started with bosonic string theory with a vanishing bulk tachyon, this will no longer be true in the region where the winding tachyons condensate. Indeed, the winding tachyons and the bulk tachyon have the same mass scale and so they will condense at the same time. Therefore, the scenario we just outlined is only meant to illustrate how such a kind of mechanism could be implemented on the worldsheet.\\

We now discuss how this worldsheet picture behaves under T-duality. The winding tachyon in the original theory is simply the bulk tachyon in the T-dual frame. We can see this from the mass formula \eqref{eq:massformula}: when $R_0\to 0$ the tower of winding states in the original theory reconstructs the bulk tachyon at level 0 in the T-dual theory. Therefore, as the winding tachyon condenses in the BoN frame, the bulk tachyon condenses in the dual frame where the circle is large. The region where the tachyons in both frames start condensing is around $r=r_0$ (as befits the stringy picture of the BoN). In the T-dual frame, this region is at strong coupling and strong curvature, as it should be, since this process cannot be described in any EFT. Nevertheless, among all of the deformations of the worldsheet that arise at this point, some will be the vertex operator \eqref{eq:tachyon} which is a momentum mode in this dual frame. This immediately signals that as one gets close to $r=r_0$, these momentum modes contribute and make it so that the solution now depends non-trivially on the dual angle $\tilde{\theta}$.

Using the potential obtained in \eqref{eq:potential}, we can even argue that there is a hypersurface defined by a line in the $\{r,\tilde \theta\}$ plane that acts as a potential barrier for all of the modes of the string, beyond which point the string modes become massive. In this way, we see that even in this dual frame, we seem to be describing a configuration ending the 26-dimensional spacetime. In this way, the configuration ending the spacetime of bosonic string theory on a circle T-dualizes to that of bosonic string theory in 26 dimensions. 
 
The fact that two configurations ending spacetime can be related by T-duality and that one of them can have a non-trivial angular dependence has already been observed in the literature in examples with a lot of supersymmetry. For instance, it was shown already in \cite{McNamara:2019rup} that the cobordism defect that kills the spin-bordism class of the circle for type IIB can be obtained from T-dualizing the O8 plane that kills the same class for type IIA on a circle. The resulting geometry described a set of 2 O7-planes at different points in the circle. Our analysis thus suggests that similar mechanisms could be at play in cases without supersymmetry. \\

We proposed a sketch of a possible stringy realization of the BoN, motivated by similar setups in the literature. It is hard to go beyond this without first studying the full sigma-model that corresponds to the BoN. It could be that the tachyonic modes play no role at all. Even if they do, one would need a full-fledged description of the condensation process described in this section beyond the classical approximation that we used in terms of barrier potentials, perhaps along the lines of \cite{Adams:2005rb}. 

\section{Final remarks}

In this note, we have taken a first step in the direction of understanding the T-duality of the bubble of nothing. Originally motivated by how T-duality plays a role in the spacetime-ending configurations predicted by the Cobordism Conjecture, we have given evidence that the T-dual of the BoN is not a spontaneous decompactification as one would naively expect from the Buscher rules. We argued that the T-dual solution depends non-trivially on the dual circle, as can be seen from the presence of winding modes on the worldsheet. 

A more precise description of the T-dual solution requires a full-fledged sigma model for the BoN. This can potentially be achieved by making it static, either by choosing a better coordinate frame or by stabilizing it with extra ingredients. We leave this for future work.  

Nevertheless, we proposed one possibility for the stringy realization of the BoN, motivated by similar setups in the literature, where closed string tachyon condensation mediates the decay of spacetime. \\

Even if a full-fledged sigma model shows that the stringy BoN is truly mediated by tachyon condensation, one would need to understand the full process of closed string tachyon condensation and go beyond the classical approximation. This is well beyond the scope of this note but it would provide tremendous insight into the decay of space time and quantum gravity away from the supersymmetric lamppost, through the cobordism conjecture. In particular, it would be interesting to further explore how geometric bubbles of nothing and tachyon condensation seem to be describing two sides of the same coin, as in \cite{FZZ,Fabinger:2000jd}.

In particular, if one could stabilize the BoN with a linear dilaton, and identify a corresponding sigma model, the case at hand would more closely resemble examples in the literature where tachyon condensation is somewhat under control, like the $\alpha'$-exact solutions of \cite{Hellerman:2006ff,Hellerman:2007fc}. These solutions describe BoNs and dimension-quenching bubbles in non-critical string theories, where the tachyon condensation process was argued to be exact in $\alpha'$, making it tractable from a worldsheet perspective. Similar bubbles can lead to time-like configurations that end spacetime, which have applications to string-inspired cosmologies \cite{Hellerman:2006hf,Hellerman:2006nx}. In fact, these bubbles have already been re-interpreted through the lens of dynamical cobordism in \cite{Angius:2022mgh}. Furthermore, the theories at the end-point of these dimension-quenching bubble processes have recently been argued to describe the near-horizon geometry of non-supersymmetric branes predicted by the cobordism conjecture \cite{Kaidi:2020jla,Kaidi:2023tqo}. Finally, in the presence of a linear dilaton background, one could potentially relate this discussion to the already-existing literature on the T-duality of cigar CFTs and its applications to 2D Euclidean black holes  \cite{Kiritsis:1991zt,Kazakov:2000pm,Hori:2001ax}. It would therefore be of great interest to see if and how generalizations of the BoN (and other examples of dynamical cobordisms to nothing) could play a similar role. \\

In the context of recent discussion around the supersymmetric lamppost, it would be relevant to contrast the T-duality of this bubble of nothing with more supersymmetric setups. In \cite{Gregory:1997te,Tong:2002rq}, the supersymmetric Taub-NUT geometry is T-dualized to an NS5 brane that is localized on the dual-circle. In this way, the T-dual of a configuration where a circle pinched off is simply an extended object in the dual frame. Moreover, in \cite{Hori:2000kt}, the T-duality of shrinking cycles has been studied in the context of Calabi-Yau compactifications, in a demonstration of mirror symmetry. It is crucial to better understand the role that supersymmetry plays in shrinking cycles and configurations that end spacetime. If one could understand the T-duality of topologically allowed BoNs in full generality, then one could imagine using T-duality to constrain the spin-defect that should arise at the point of pinch-off in cases where they are topologically obstructed. This could then lead to the characterization of unknown non-supersymmetric defects in string theory.\\

Relatedly, it would be interesting to generalize these results to more general spacetime-ending configurations. Indeed, the BoN constitutes a simple example of the more general notion of a dynamical cobordism to nothing \cite{Buratti:2021fiv,Angius:2022aeq}. The framework of dynamical cobordism works from the perspective of the lower-dimensional EFT obtained when compactifying on the compact space that is pinching off. The locus of this pinch-off is the position of a so-called End-of-the-World (ETW) brane: a codimension-1 surface where the EFT breaks down. These ETW brane solutions were described in \cite{Angius:2022aeq} as the back-reaction on supergravity of the boundary predicted by the cobordism conjecture. It turns out that close enough to the singularity, these solutions all explore infinite distance in field space. In some cases, the infinite distance limit explored corresponded to a decompactification limit instead of a pinching off of the compact geometry. This was difficult to match to any notion of cobordism, and was one of the motivations for this work.

Once compactified on the circle, the BoN and its dual both fit in the framework of ETW brane solutions. It would therefore be of relevance to understand how T-duality acts on these solutions and gain insight into the stringy physics that arise near ETW banes from the worldsheet perspective. Indeed, since these configurations are generically singular and/or at strong coupling, we expect many corrections to become relevant near their core. If one could better understand such effects, one could potentially identify \textit{which} of the ETW brane solutions actually describe a cobordism to nothing in the UV. In particular, in examples of ETW branes in superstring theories, one can hope to shed light on this question by using other known string dualities to probe the strongly coupled core.  \\

We hope to come back to these questions in the near future.

\section*{Acknowledgements}
We are very grateful to Miguel Montero and Angel Uranga for their invaluable contributions through extensive discussions and thoughtful comments on the manuscript. We are also pleased to thank Roberta Angius, Ginevra Buratti, José Calderón-Infante and Jesús Huertas for useful discussions and for collaborations on related topics. We also thank Alexander Bernal, Pau Garcia Romeu, Jacob McNamara, Gonzalo Morras, Lorenzo Paoloni, Ignacio Ruiz, Irene Valenzuela and Yoav Zigdon for helpful discussions and comments. The work of MD is supported by the FPI grant no. FPI SEV-2016-0597-19-3 from Spanish National Research Agency from the Ministry of Science and Innovation. MD acknowledges the support of the grants CEX2020-001007-S and PID2021-123017NB-I00, funded by MCIN/AEI/10.13039/501100011033 and by ERDF A way of making Europe.

\bibliographystyle{unsrt}
\bibliography{this.bib}

\end{document}